\documentclass[]{spie}           
\usepackage[dvips]{graphicx}

\DeclareTextSymbolDefault{\micro}{TS1}
\DeclareTextSymbol{\micro}{TS1}{181} 
\DeclareFontEncoding{TS1}{}{}
\DeclareFontSubstitution{TS1}{cmr}{m}{n}

\title{10-{\Large \textbf{\micro}}m wavefront spatial filtering: \\
       first results with chalcogenide fibers} 

\author{Pascal Bord\'e\supit{a}, Guy Perrin\supit{a},
Thanh Nguyen\supit{b}, Anne Amy-Klein\supit{b}, Christophe Daussy\supit{b}, \\
Pierre-Ivan Raynal\supit{c}, Alain L\'eger\supit{c}
and Gwena\"el Maz\'e\supit{d}
\skiplinehalf
\supit{a}LESIA, Observatoire de Paris, Meudon, France \\
\supit{b}Laboratoire de Physique des Lasers, Universit\'e Paris-Nord,
Villetaneuse, France \\
\supit{c}Institut d'Astrophysique Spatiale, Universit\'e Paris-Sud, Orsay,
France \\
\supit{d}Le Verre Fluor\'e, Campus de Ker Lann, Bruz, France
}

\authorinfo{Further author information: (Send correspondence to P.B.) \\
P.B.: E-mail: Pascal.Borde@obspm.fr \\
Copyright 2002 Society of Photo-Optical Instrumentation Engineers. \\
This paper will be published in the proceedings of the conference
\emph{Interferometry for Optical Astronomy II}, part of SPIE's Astronomical
Telescopes \& Instrumentation, 22--28 August 2002, Waikoloa, HI and is made
available as an electronic preprint with permission of SPIE. One print or
electronic copy may be made for personal use only. Systematic or multiple
reproduction, distribution to multiple locations via electronic or other means,
duplication of any material in this paper for a fee or for commercial purposes,
or modification of the content of the paper are prohibited.
}
 
\begin{document} 
  \maketitle 

%
\begin{abstract}
Wavefront cleaning by single-mode fibers has proved to be efficient in
optical-infrared interferometry to improve calibration quality.  For instance,
the FLUOR instrument has demonstrated the capability of fluoride glass
single-mode fibers in this respect in the K and L bands. New interferometric
instruments developped for the mid-infrared require the same capability for the
8--12~{\micro}m range. We have initiated a program to develop single-mode
fibers in the prospect of the VLTI mid-infrared instrument MIDI and of the
ESA/DARWIN and NASA/TPF missions that require excellent wavefront quality.
In order to characterize the performances of chalcogenide fibers we are
developping, we have set up an experiment to measure the far-field pattern
radiated at 10~{\micro}m. In this paper, we report the first and promising
results obtained with this new component.
\end{abstract}


%
\section{INTRODUCTION} \label{sec:intro}
Whatever the cause of corrugations on interfering wavefronts, defects on the
mirrors or effect of the turbulent atmosphere, they result in a loss in the
coherence factor measured by an interferometer. Spatial filtering by a pinhole
overcomes this problem by removing the high spatial frequencies from the
wavefronts. However, the pinhole has to be designed for a specific wavelength
and does not provide a good correction of low spatial frequency defects.
An alternative consists in modal filtering by a single-mode waveguide: the
shape of the wavefronts that have been guided into this device only depends on
the characteristics of the guide (choice of materials and geometry); this way,
two beams can be recombined with a full interferometric efficiency. This
property holds for wavelengths above the cutoff wavelength of the guide, as the
only mode to propagate is then the fundamental one.

Since 1995, the FLUOR\cite{Foresto98} instrument has demonstrated the use of
fluoride single-mode fibers to measure high precision coherence factors on many
stars in the near infrared (K and L bands). It is now followed by the VLTI
near-infrared instruments VINCI\cite{Kervella00} (K band) and
AMBER\cite{Petrov00} (H and K bands). Unfortunately, no single-mode fiber is
available nowadays for the mid-infrared. This is an issue for the VLTI
10-{\micro}m instrument MIDI\cite{Leinert00} (N band) and for its ambitious
followers on the ground like GENIE\cite{Gondoin02}, or in space like
DARWIN\cite{Leger96} that aims at performing the spectroscopy of exoplanets in
the 8--20 {\micro}m range. In this prospect, we have undertaken an activity of
research and development on single-mode fibers for the mid-infrared, with the
partnership of Le Verre Fluor\'e, a leading French company in this sector.
In this paper, we report the promising results obtained with chalcogenide
fibers in terms of transmission, wavelength range and modal filtering.

In Sect.~\ref{sec:proto}, we review the fundamental properties of single-mode
fibers and detail the characteristics of the prototype manufactured by Le
Verre Fluor\'e. Section~\ref{sec:meas} first describes the experiment that has
been set up to test the filtering capability of the fibers, and then discusses
the results of our measurements.
%
%
\section{PROTOTYPE MANUFACTURING} \label{sec:proto}
%
%
\subsection{Single-mode fibers} \label{sub:fibers}
An optical fiber becomes single-mode when it is used at a wavelength longer
than the cutoff wavelength
\begin{equation} \label{eq:lbda}
\lambda_\mathrm{c} = \frac{2\pi \, a \; \mathrm{NA}}{2.405}
\quad \mathrm{with} \quad \mathrm{NA} = \sqrt{n^2_\mathrm{c}-n^2_\mathrm{g}},
\end{equation}
where $a$ is the core radius, NA is the numerical aperture, $n^2_\mathrm{c}$
and $n^2_\mathrm{g}$ are respectively the core and the cladding refractive
indices\cite{Neumann88}. In this situation, the only guided mode is the
fundamental one, denoted LP$_{01}$, whose shape is almost Gaussian. All the
energy injected into the fiber will either excite this mode or propagate into
the cladding. If the fiber is long enough, the cladding modes are completely
attenuated and do not come out.
%
%
\subsection{Selected materials} \label{sub:materials}
For this applied research, Le Verre Fluor\'e was funded by the French military
administration, la Direction G\'en\'erale de l'Armement (DGA), and associated
to three research laboratories for the full characterization of the produced
fibers: the Laboratoire de Physique des Lasers (LPL), the Institut
d'Astrophysique Spatiale (IAS), and the D\'epartement de Recherches Spatiales
(now LESIA). The goal assigned by the one-year contract (dec.~1999--dec.~2000)
was to manufacture a single-mode waveguide for the 8--12~{\micro}m range with
losses less than 3~dB/cm.

Two solutions were selected by Le Verre Fluor\'e to achieve this goal, as
transparent materials in the mid-infrared suitable for fiber manufacturing are
either
\begin{description}
\item{(a)} silver halides, e.g. AgCl or AgBr, or
\item{(b)} chalcogenides, e.g. As$_x$Se$_y$Te$_z$.
\end{description}
In order to obtain a specific refractive index, different halides are mixed
together in case~(a), whereas the proportions x, y and z are varied in
case~(b). Silver halides have the appealing property of being transparent up to
30~{\micro}m with transmission losses that can be as low as 10$^{-3}$~dB/km.
However, they are polycristals and as such are turned into fibers by extrusion
(the material is pushed through a hole), a process difficult to master. On the
other hand, chalcogenides are glasses, so fibers can be manufactured by
drawing, but they become opaque beyond 16~{\micro}m. In the prospect of the
short-term DGA contract, the second solution appeared easier and was prefered.
Nevertheless, silver halides are certainly the material that will be needed in
the future for ground-based applications in the Q band (17--25~{\micro}m),
or for DARWIN in space at the longest wavelengths of the spectrum.
%
%
\subsection{First prototype characteristics} \label{sub:charac}
\begin{figure}[htbp]
\begin{center}
\begin{tabular}{c}
\includegraphics[width=12cm]{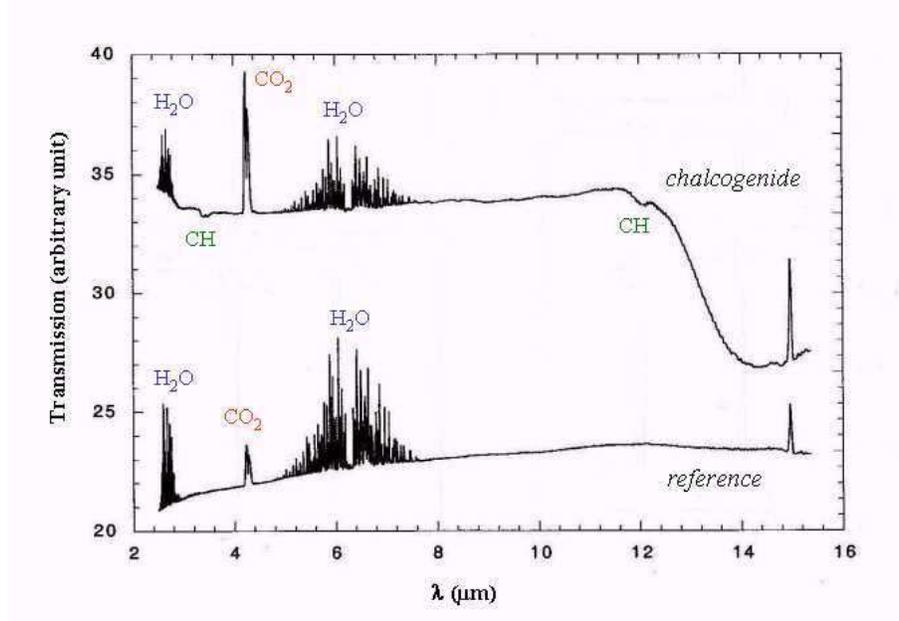}
\end{tabular}
\end{center}
\caption{ \label{fig:LURE} Transmission spectrum of the chalcogenide glass.
Only the CH lines belongs to the material's spectrum, not those seen in
emission (H$_2$O, CO$_2$) that are features of the reference spectrum and
originates from the air in the spectrometer.}
\end{figure}

The characteristics of the first fiber prototype are listed in
Table~\ref{tab:charac}.
\begin{table}[h]
\caption{Characteristics of the first fiber prototype manufactured by Le Verre
Fluor\'e.} 
\label{tab:charac}
\begin{center}       
\begin{tabular}{|l|l|}
\hline
\rule[-1ex]{0pt}{3.5ex}  Material            & As$_2$Se$_3$/GeSeTe$_{1.4}$ \\
\hline
\rule[-1ex]{0pt}{3.5ex}  Core diameter       & 40~{\micro}m \\
\hline
\rule[-1ex]{0pt}{3.5ex}  Cladding diameter   & 210~{\micro}m \\
\hline
\rule[-1ex]{0pt}{3.5ex}  Numerical aperture  & 0.15 \\
\hline
\rule[-1ex]{0pt}{3.5ex}  Cutoff wavelength   & 8.1~{\micro}m \\
\hline
\rule[-1ex]{0pt}{3.5ex}  Length              & 8~cm \\
\hline
\rule[-1ex]{0pt}{3.5ex}  Transmission losses & $\approx 0.2$~dB/cm \\
\hline 
\end{tabular}
\end{center}
\end{table}
The transmission losses were determined by Le Verre Fluor\'e using the cut-back
method: the throughput of a sample is measured twice, before and after
shortening one end; the difference in throughput leads to the transmission
losses given the length of the missing piece. The losses amount to
$\approx 30\%$ at 10~{\micro}m for our 8-cm prototype. Besides, the cutoff
wavelength leaves a signature in the spectral loss curve: the transition
between two to one guided mode results in a downward slope. The cutoff
wavelength was thus estimated to 8.1~{\micro}m. A high-resolution transmission
spectrum of the material itself was also recorded with the FTS microscope of the
LURE laboratory in Orsay (Fig.~\ref{fig:LURE}). This spectrum shows a flat 
transmission until 12~{\micro}m and no impurity except for a very little
quantity of CH that will be eliminated when the next fiber is drawn.

Although these first results seem encouraging, the ultimate test remains to
inject a corrugated wavefront in the fiber and to check the field profile at
the end. In the next section, we describe this experiment and its results.
%
%
\section{FAR-FIELD MEASUREMENTS} \label{sec:meas}
%
%
\subsection{LPL testbed} \label{sub:testbed}
\begin{figure}[htbp]
\begin{center}
\begin{tabular}{c}
\includegraphics[width=12cm]{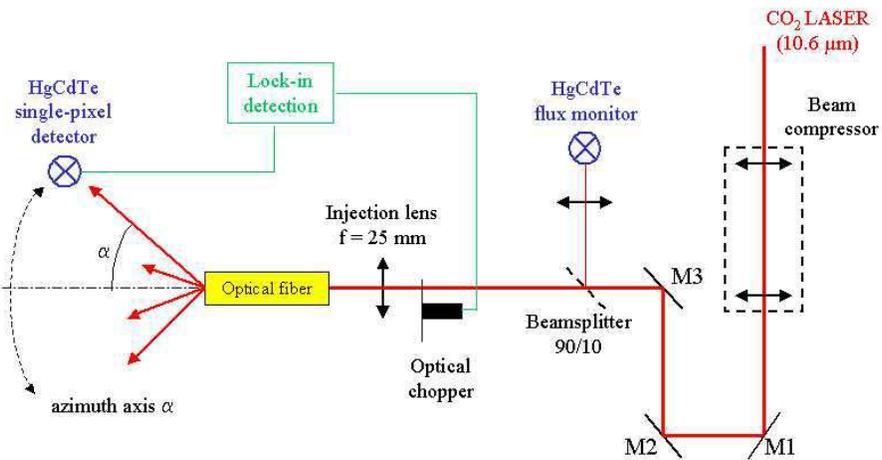}
\end{tabular}
\end{center}
\caption{ \label{fig:layout} Layout of the testbed hosted by the Laboratoire
de Physique des Lasers.}
\end{figure}

\begin{figure}[htbp]
\begin{center}
\begin{tabular}{c}
\includegraphics[width=12cm]{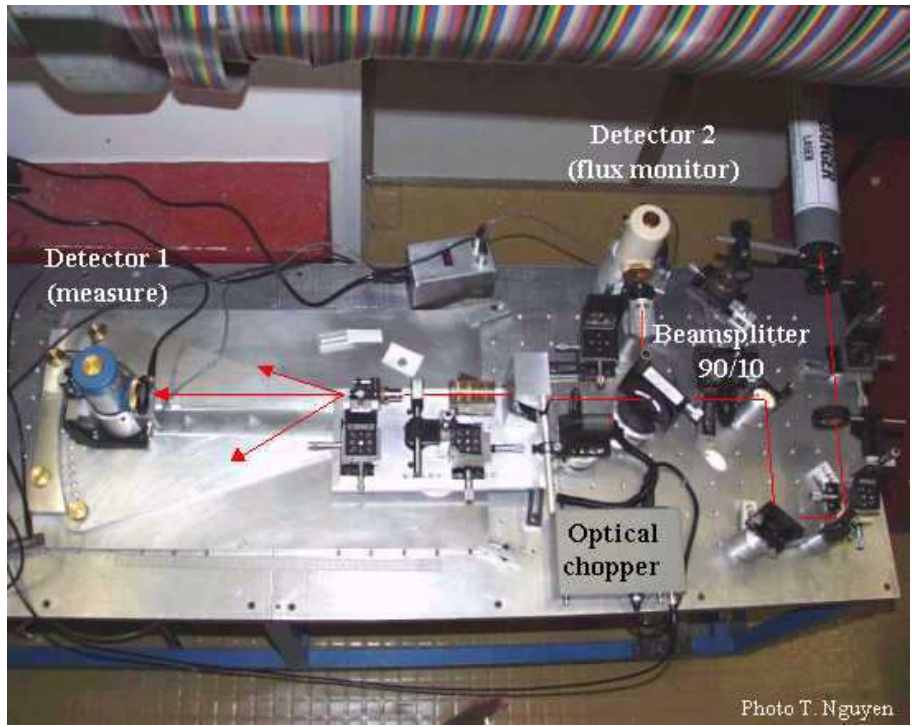}
\end{tabular}
\end{center}
\caption{ \label{fig:photo} Picture of the testbed hosted by the Laboratoire
de Physique des Lasers.}
\end{figure}

Single-mode fiber theory\cite{Neumann88} establishes that the fiber end
radiates a free diverging wave beam, the fundamental fiber mode becoming a
fundamental Gaussian mode with a good approximation. The waist of the radiated
Gaussian beam is located on the fiber end and has a radius equal to the radius
of the fiber core. We have set up an experiment
(Figs.~\ref{fig:layout}--\ref{fig:photo}) to check the quality of the modal
filtering by measuring the far-field radiation pattern of the fiber
(experiments of this kind have already been reported in the
literature\cite{Hotate79}).

The testbed is hosted by the Laboratoire de Physique des Lasers, a laboratory
that operates several highly stabilized CO$_2$ lasers designed for metrology
applications. A fraction of a 10.6-{\micro}m laser beam, primarily dedicated to
another experiment, is folded in the direction of our optical table. The beam
is first recollimated by an afocal telescope. Part of the light is then sent to
a flux monitor (HgCdTe monopixel detector), whereas the main beam is adjusted
in height by a periscope before being focused onto the fiber core by an
injection lens. The light propagates along the 8-cm fiber and is radiated
freely in space. If all the modes but the fundamental have been filtered out,
the energy is expected to be radiated within in a cone of about
$2\theta = 2\arctan [\lambda/(\pi \, w)] \approx 20^\circ$, since the waist
radius is $w = a/2 = 20$~{\micro}m.

The far-field pattern is recorded by moving a single-pixel HgCdTe detector
in the portion of space in front of the fiber. Because of the high background
emissivity at 10.6~{\micro}m, a lock-in detection system is mandatory.
The detector can be positioned at three different distances from the fiber end:
10, 20 or 30~cm and is adjustable in height. The exploration of the 3D
far-field pattern ($\pm 20^\circ$ in azimuth, $\pm 5^\circ$ in height) is done
manually, and is completed in about one hour and a half for a resolution of
$1^\circ$.

\begin{figure}[htbp]
\begin{center}
\begin{tabular}{c}
\includegraphics[width=16cm]{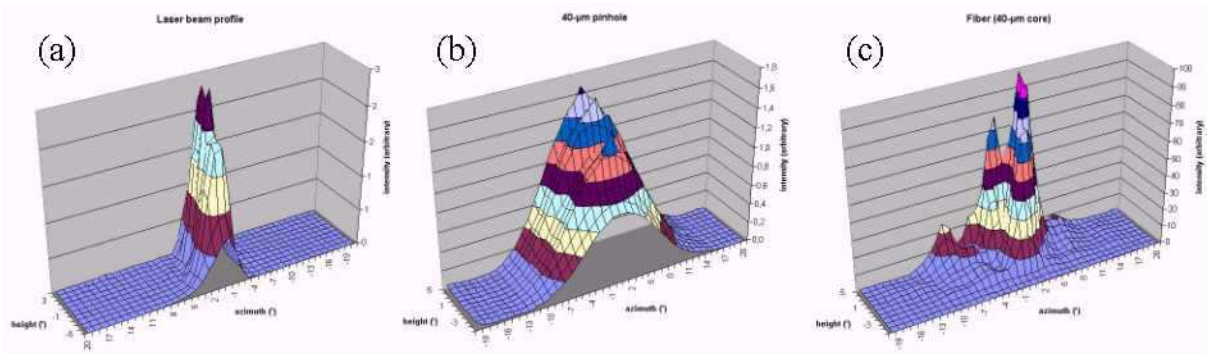}
\end{tabular}
\end{center}
\caption{ \label{fig:3D-prof} Far-field radiation patterns: (a) laser beam,
(b) 40-{\micro}m pinhole, and (c) 40-{\micro}m core chalcogenide fiber.}
\end{figure}

We have first measured the laser beam profile by replacing the fiber by a
convergent lens to produce a divergent beam in a cone opened enough to be
sampled by our system: this profile (Fig.~\ref{fig:3D-prof}a) is not Gaussian
as the laser is not single-mode. In a second step, the experimental procedure
is validated by recording the far-field radiation pattern of a 40-{\micro}m
pinhole: the profile (Fig.~\ref{fig:3D-prof}b) has the expected Gaussian shape
and width.
%
%
\subsection{First far-field profiles} \label{sub:first}
At our great disappointment, the intensity profile radiated by the fiber was
not found to be Gaussian at all (Fig.~\ref{fig:3D-prof}c). Three modes can be
seen in the portion of space that was explored. Only the radius of the whole
bunch of modes is compatible with a diffraction by the fiber core, whereas
their individual radii correspond more closely to a diffraction by the
cladding. This indicates that the cladding modes are not
filtered out. A first explanation could be that the fiber sample is simply not
long enough to eliminate the unguided light. However, a previous experiment
with a 1-mm long fluoride glass fiber designed for 2~{\micro}m and used at
10~{\micro}m have shown better results\cite{Perrin00}. The correct explanation
was found by taking into account the protective layer surrounding the cladding.
This extra layer happens to have a refractive index inferior to that of the
cladding itself. This results in a second guiding structure, concentric to the
core-cladding pair, that causes the persistent cladding modes.
As a consequence, the transmission figure of Table~\ref{tab:charac} is
probably overestimated as it includes unwanted energy. The problem could be
solved if another material were chosen for the protective layer, but at this
date, the DGA contract came to an end. Nevertheless, some research could be
pursued on a collaborative basis with Le Verre Fluor\'e.
%
%
\subsection{Second generation samples} \label{sub:second}
\begin{figure}[htbp]
\begin{center}
\begin{tabular}{c}
\includegraphics[width=16cm]{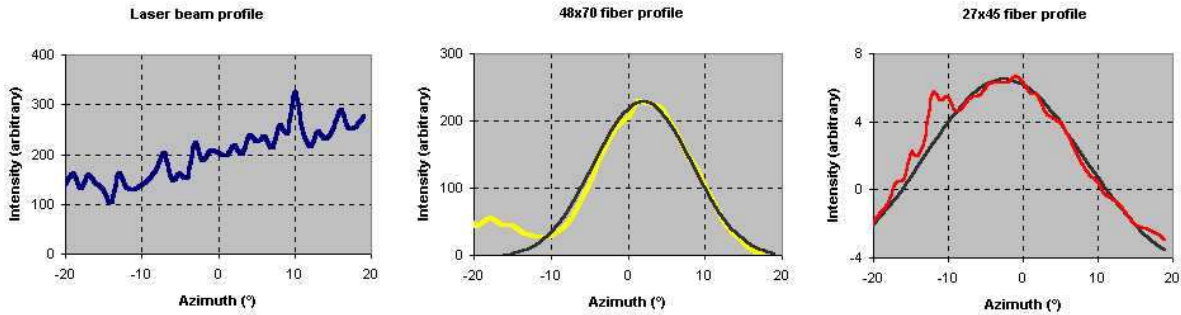}
\end{tabular}
\end{center}
\caption{ \label{fig:1D-prof} Far-field radiation patterns (from left to
right): laser beam, 48~{\micro}m/70~{\micro}m and 27~{\micro}m/45~{\micro}m
fibers. Best-fit Gaussian curves in black are superimposed to both fiber
profiles.}
\end{figure}

Because of the lack of fundings and despite the lessons learnt from the first
campaign of measurements, the fibers produced in the frame of the DGA
contract had to be re-used.
A second generation of fibers were obtained by removing the extra layer and
part of the cladding of the first generation by chemical stripping. Thus, the
cladding is studded with diffusive centers favoring the elimination of cladding
modes. Besides, the fibers were coated with a new protective enveloppe in
lead, a 10~{\micro}m absorbent. The two new components are shorter, 4-mm long,
with thinner claddings: the core/cladding dimensions are
48~{\micro}m/70~{\micro}m and 27~{\micro}m/45~{\micro}m respectively.
Figures~\ref{fig:1D-prof} display the 1D far-field radiation patterns of the
laser beam and of both fibers that were recorded during a second measurement
campaign. A best-fit Gaussian curve has been superimposed for comparison.
Although there is still a small second lobe in one case
(48~{\micro}m/70~{\micro}m), the profiles appear to be much closer to what we
are seeking. However, their width is not completely consistent with the
prediction of the standard diffraction theory. We think the cladding is so thin
that it can no longer be considered as infinite as it is usually assumed in the
standard fiber theory. The cladding will be thicker when the industrial process
is ready. A close examination of the fibers with a binocular microscope has
revealed geometrical defects and a small asymmetry of the fiber cores that are
lilkely to be responsible for the high frequency defects and the small second
lobe. This has been recently improved by Le Verre Fluor\'e, still using the
fibers originally drawn, and we plan to put under test these new components
this fall.
%
%
\section{CONCLUSION} \label{sec:concl}
We have presented the characteristics of chalocogenide fibers we are
developping to perform modal filtering in the mid-infrared. We have reported
promising first results with these components. Because of our present lack
of funding, we still work on the fibers that were drawn in the first place.
As they could recently be improved nonetheless, and are now ready for testing,
we will start a new measurement campaign next fall. If these fibers are
successfully tested, they could be integrated in MIDI next year. The following
step would be then to develop silver halide fibers for modal filtering up to
20~{\micro}m, i.e. through the full wavelength range of DARWIN.
%
%

\end{document}